\documentclass[12pt]{article}
\usepackage[T2A]{fontenc}
\usepackage[cp1251]{inputenc}
\usepackage{hyperref}
\usepackage{a4wide}
\usepackage[english]{babel}
\usepackage{amsmath,latexsym,amsthm,amsfonts}
\usepackage{amssymb,amscd}
\usepackage{graphicx}

\begin{document}
\title{ Currency Derivatives Pricing for Markov-modulated Merton Jump-diffusion Spot Forex Rate}
\author{Anatoliy Swishchuk$^1$, Maksym Tertychnyi $^2$,   Winsor Hoang $^3$}
\date{}
\maketitle
\begin{footnotesize}
\begin{tabbing}
$^1$ \= Department of Mathematics and Statistics, University of Calgary, Canada\\
\> {\em aswish@ucalgary.ca}\\
$^2$ \> Department of Mathematics and Statistics, University of Calgary, Canada\\
\>{\em mtertych@ucalgary.ca, maksym.tertychnyi@gmail.com}\\
$^3$ \=  CTS Forex, Canada\\
\> {\em winsorhoang@ctsforex.com}\\
\end{tabbing}
\end{footnotesize}

\begin{abstract}
We derived similar  to Bo {\it et al.} (2010) results but in the case when the dynamics of the  FX rate is driven by a general Merton jump-diffusion  process.  The main results of our paper are as follows: 1) formulas for the Esscher transform parameters which ensure that the martingale condition for the discounted foreign exchange rate is a martingale for a general Merton jump--diffusion process are derived; using the values of these parameters we proceeded to a risk-neural measure and provide new formulas for the distribution of jumps, the mean jump size, and  the Poisson process intensity with respect  to the measure;  pricing formulas for European call foreign exchange options have been given as well; 2) obtained formulas are applied to the case of  the exponential processes; 3) numerical simulations of European call foreign exchange option prices for  different parameters are also provided; 4) codes for Matlab  functions used in numerical simulations of option prices are given.
\end{abstract}
\thispagestyle{empty}



\section{\bf Introduction}

The existing academic literature on the pricing of foreign currency options could be divided into two
categories: 1) both domestic and foreign interest rates were assumed to be constant whereas
the spot exchange rate is assumed to be stochastic (see, e.g., Jarrow et al. (1981, \cite{b15-2}); 2)  models for pricing foreign currency options incorporate stochastic interest rates, and are based on Merton's 1973, \cite{b22}) stochastic interest rate model for pricing equity options (see, e.g., Grabbe (1983, \cite{b16}),
Adams et al. (1987, \cite{b1}). In both cases, this pricing approach did not integrate a full term structure
model into the valuation framework. To our  knowledge, Amin et al. (1991, \cite{b3}) were the first  to start discussing and building a general
framework to price contingent claims on foreign currencies under stochastic interest rates using the
Heath et al. (1987) model of term structure. Melino et al. (1991, \cite{b20}) examined the foreign exchange rate
process, (under a deterministic interest rate), underlying  observed option prices and Rumsey (1991, \cite{b28})
considered cross-currency options. Mikkelsen (2001, \cite{b24}) investigated by simulation cross-currency options using market
models of interest rates and deterministic volatilities for spot exchange rates. Schlogl
(2002, \cite{b29}) extended market models to a cross-currency framework. Piterbarg (2005, \cite{b26}) developed a model
for cross-currency derivatives such as PRDC swaps with calibration for currency options; he used
neither market models nor stochastic volatility models. In Garman et al. (1983, \cite{b9}) and Grabbe (1983, \cite{b16}),
foreign exchange option valuation formulas were derived under the assumption that the exchange rate
follows a diffusion process with continuous sample paths.
Takahashi et al. (2006, \cite{b32}) proposed a new approximation formula for the valuation of currency options
 using  jump-diffusion stochastic volatility processes for spot exchange rates in a stochastic interest rates
environment. In particular, they applied the  market models developed by Brace et al. (1998), Jamshidian
(1997, \cite{b15}) and Miltersen et al. (1997, \cite{b23}) to model the term structure of interest rates. Also, Ahn et al. (2007, \cite{b2})
derived explicit formulas for European foreign exchange call and put options values when the
exchange rate dynamics are governed by jump-diffusion processes.
Hamilton (1988) was the first to investigate the term structure of interest rates by rational expectations
econometric analysis of changes in regime. Goutte et al. (2011, \cite{b10}) studied foreign exchange
rates using a modified Cox-Ingersoll-Ross model under a Hamilton Markov regime switching
framework. Zhou et al. (2012, \cite{b33}) considered an accessible implementation of interest rate models with
regime-switching. Siu et al (2008, \cite{b31}) considered pricing currency options under a two-factor Markov modulated
stochastic volatility model.  Swishchuk and Elliott   applied hidden Markov models for pricing options in \cite{b31-1}.  Bo et al. (2010, \cite{b7}) discussed a  Markov-modulated jump-diffusion,
(modeled by a compound Poisson process), for currency option pricing.
We note that currency derivatives for domestic and foreign equity markets and for the exchange rate
between the domestic currency and a fixed foreign currency with constant interest rates were discussed
in Bjork (1998, \cite{b6}). We also mention that currency conversion for forward and swap prices with constant
domestic and foreign interest rates were discussed in Benth et al. (2008, \cite{b5}).

In this article we generalize results in \cite{b7} on a case when dynamics of  FX rate is driven by general Merton jump-diffusion process (\cite{b21}). Main results of our research are as follows:

1) In section 2 we generalize formulas in \cite{b7} for Esscher transform parameters assuring that martingale condition for discounted foreign exchange rate is a martingale for a general Merton jump-diffusion  process (see \eqref{4.74}). Using these values of parameters  (see \eqref{4.82}, \eqref{4.83}) we proceed to a risk-neural measure and provide new formulas for the distribution of jumps (\eqref{4.80}), the mean jump size (see \eqref{4.67}), and  the Poisson process intensity with respect  to this measure (see \eqref{4.66}).  At the end of section 2
 pricing formulas for a European call foreign exchange option  are  given (They are similar to those in \cite{b7}, but the mean jump size and the Poisson process intensity with respect  to the new risk-neutral measure are different).

2) In section 3 we apply formulas \eqref{4.65}-\eqref{4.67}, \eqref{4.82}-\eqref{4.83} to a particular case of the exponential distribution (see \eqref{4.84}) of jumps (see \eqref{4.87}-\eqref{4.89}).

3) In section 4 we provide numerical simulations of European call foreign exchange option prices for  different parameters:  $S/K$, where $S$ is the initial spot FX  rate, $K$ is the strike FX rate for a  maturity time $T$.

Appendix contains the codes for Matlab  functions used in numerical simulations of option prices.

\section{Currency option pricing  for  Merton jump-diffusion processes}

Let $(\Omega, \mathcal{F}, \textbf{P})$ be a complete probability space with a  probability measure $\textbf{P}$. Consider a continuous-time, finite-state
Markov chain $\xi= \{\xi_t\}_{0\leq t \leq T}$  on
$(\Omega, \mathcal{F}, \textbf{P})$ with a state space $\mathcal{S}$, the set of unit vectors $(e_1,\cdots, e_n)\in \mathbb{R}^n$ with a rate matrix $\Pi$\footnote{In our numerical simulations we consider three-state Markov chain and calculate elements in $\Pi$ using Forex market EURO/USD currency pair}.
The dynamics of the chain are given by:
 \begin{equation}\label{3.11-1}
 \xi_t=\xi_0+\int_0^t \Pi \xi_u du+M_t \in \mathbb{R}^n,
\end{equation}
where  $M = \{M_t, t \geq 0\}$ is a $\mathbb{R}^n$-valued martingale with respect to $(\mathcal{F}_t^\xi)_{0 \leq t \leq T}$, the $\textbf{P}$-augmentation of the natural filtration $(\mathcal{F}_t)_{0 \leq t \leq T}$, generated by the Markov chain $\xi$. Consider  a Markov-modulated Merton jump-diffusion which models the dynamics of the spot FX rate, given by the following stochastic differential equation (in the sequel SDE, see \cite{b7}):
\begin{equation}\label{3.11}
dS_t= S_{t_-}\left(\mu_t dt+\sigma_t d W_t+(Z_{t_-}-1)dN_t\right), \; Z_t>0.
\end{equation}
Here $\mu_t$  is drift parameter; $W_t$ is a  Brownian motion, $\sigma_t$ is the  volatility; $N_t$ is a  Poisson Process with intensity $\lambda_t$,  $Z_{t_-}-1$ is the amplitude of the jumps, given the  jump arrival time. The distribution of $Z_t$ has a density $\nu(x), x\in \mathbb{R}$.
The parameters $\mu_t$,\;$\sigma_t$,\;$\lambda_t$ are modeled using the finite state Markov chain:
\begin{align}\label{3.11-1}
&\mu_t:=<\mu, \xi_t>, \; \mu \in \mathbb{R}^n_+;\notag \\
&\sigma_t:=<\sigma, \xi_t>, \; \sigma \in \mathbb{R}^n_+;\notag\\
&\lambda_t:=<\lambda, \xi_t>, \; \lambda \in \mathbb{R}^n_+.
\end{align}
The solution of \eqref{3.11} is  $S_t=S_0 e^{L_t}$, (where $S_0$ is the spot FX rate at time $t=0$). Here  $L_t$ is given by the formula:
\begin{equation}\label{3.12}
L_t=\int_0^t(\mu_s-1/2 \sigma_s^2) ds+\int_0^t \sigma_s dW_s+\int_0^t \log Z_{s_{-}}d N_s.
\end{equation}

Note, that for the most of well-known distributions (normal ,  exponential distribution of $Z_t$, etc) $L_t$ is not a L$\acute{\textnormal{e}}$vy process (see definition of L$\acute{\textnormal{e}}$vy process in \cite{b25}, the condition \textbf{L3}), since $ \log Z_{t_{-}}\rightarrow -\infty$ for small $Z_t$, but probability of jumps with even  0 amplitude is a positive constant, depending on a type of distribution. We  call the process \eqref{3.12} as Merton jump-diffusion process (see \cite{b21}, section 2,  formulas 2, 3)

There is more than one equivalent martingale measure for this market driven by a Markov-modulated jump-diffusion model.  We shall
define the regime-switching generalized Esscher transform to determine a specific equivalent martingale measure.

Using Ito's formula we can derive a stochastic differential equation for the discounted spot FX rate.
To define the discounted spot FX rate   we need to introduce domestic and foreign riskless interest rates for bonds in the domestic and foreign currency.

The domestic and foreign interest rates $(r^d_t)_{0\leq t\leq T}$,
$(r^f_t)_{0\leq t\leq T}$ are defined using the Markov
chain $(\xi_t)_{0\leq t\leq T}$ (see \cite{b7}):
$$ r^d_t=\langle r^d, \xi_t\rangle, r^d \in \mathbb{R}_+^n,$$
$$ r^f_t=\langle r^f, \xi_t\rangle, r^f \in \mathbb{R}_+^n.$$
The discounted spot FX rate is:
\begin{equation}\label{3.13}
 S^d_t=\textnormal{exp}\left(\int_0^t (r^d_s-r^f_s) ds\right)S_t, \quad
 0\leq t\leq T.
\end{equation}
Using \eqref{3.13}, the differentiation formula, see Elliott et al. (1982, \cite{b11-1}) and the   stochastic differential equation for the spot FX rate \eqref{3.11} we find the stochastic differential equation for the discounted discounted spot FX rate:
\begin{equation}\label{3.14}
d S_{
t_-}^d= S_{t_-}^d(r^d_t-r^f_t+\mu_t) dt+
S_{t_-}^d\sigma_t d W_t
  +S_{t_-}^d(Z_{t_-}-1) d N_t.
\end{equation}

To derive the main results consider the log spot FX rate

$$Y_t= \log\left(\frac{S_t}{S_0}\right)$$
Using the differentiation formula:

$$Y_t=C_t+J_t,$$
where $C_t, J_t$ are the  continuous and diffusion part of $Y_t$. They are given  in \eqref{3.15}, \eqref{3.16}:

\begin{equation}\label{3.15}
  C_t=\int_0^t\left (r^d_s-r^f_s+\mu_s
  \right)ds+\int_0^t \sigma_s dW_s,
\end{equation}

\begin{equation}\label{3.16}
 J_t=\int_0^t \log Z_{s_-}dN_s.
\end{equation}

 Let  $(\mathcal{F}_t^Y)_{0 \leq t \leq T}$ denote the $\textbf{P}$-augmentation of the natural filtration $(\mathcal{F}_t)_{0 \leq t \leq T}$, generated by  $Y$. For each $t \in [0, T]$ set $\mathcal{H}_t=\mathcal{F}_t^Y\vee \mathcal{F}_t^\xi$. Let us also define two families of regime switching parameters

$(\theta_s^c)_{0\leq t\leq T}$, $(\theta_s^J)_{0\leq t\leq T}$:
$\theta_t^m=<\theta^m, \xi_t>$, $\theta^m=(\theta_1^m,..., \theta_n^m)\subset \mathbb{R}^n$, $m=\{c, J\}$.

Define a random Esscher transform $\textbf{Q}^{\theta^c, \theta^J}\sim \textbf{P}$ on $\mathcal{H}_t$ using these  families of parameters $(\theta_s^c)_{0\leq t\leq T}$, $(\theta_s^J)_{0\leq t\leq T}$ (see \cite{b7}, \cite{b11}, \cite{b12} for details):

\begin{equation}\label{4.1}
L_t^{\theta^c, \theta^J}=\frac{d \textbf{Q}^{\theta^c, \theta^J}}{d
\textbf{P}}\biggl|_{\mathcal{H}_t}=:
\end{equation}
$$
  \frac{\textnormal{exp} \left(\int_0^t \theta_s^c dC_s +\int_0^t \theta_{s_-}^J d J_s\right )}
  {\mathbb{E}\left [\textnormal{exp} \biggl(\int_0^t \theta_s^c dC_s +\int_0^t \theta_{s_-}^J d J_s\right )\biggl | \mathcal{F}_t^\xi\biggl]}.
$$

The explicit formula for the density $L_t^{\theta^c, \theta^J}$ of the Esscher transform is given in the following Theorem. A similar statement is proven for the log-normal distribution in \cite{b7}. The formula below can be obtained by another approach, considered by Elliott and Osakwe (\cite{b11-2}).

\textbf{Theorem 2.1.} \textit{For $0\leq t \leq T$ density $L_t^{\theta^c, \theta^J}$ of Esscher transform defined in \eqref{4.1} is given by }
\begin{equation}\label{4.59}
L_t^{\theta^c, \theta^J}=\textnormal{exp}\left(\int_0^t \theta_s^c
\sigma_s dW_s -1/2 \int_0^t (\theta_s^c \sigma_s)^2
ds\right)\times
\end{equation}
$$\textnormal{exp}\left(\int_0^t \theta_{s_-}^J \log Z_{s_-}d N_s-\int_0^t \lambda_s\biggl(\int_{\mathbb{R_+}}x^{\theta_s^J }\nu(dx)-1\right)ds\biggl).$$

\textit{In addition,the random Esscher transform density
$L_t^{\theta^c,\theta^J}$ (see \eqref{4.1}, \eqref{4.59}) is an exponential $(\mathcal{H}_t)_{0\leq t\leq T}$ martingale and admits the following SDE}

\begin{equation}\label{4.60}
  \frac{d L_t^{\theta^c,
 \theta^J}}{L_{t_-}^{\theta^c,
 \theta^J}}=\theta_t^c \sigma_t dW_t+(Z_{t_-}^{\theta_{t_-}^J}
-1)d N_t-\lambda_t \left(\int_{\mathbb{R_+}}x^{\theta_t^J }\nu(dx)-1\right)dt.
\end{equation}
\textbf{Proof Theorem 2.1.} The compound Poisson Process, driving jumps $\sum_0^{N_t}(Z_{i}-1)$, and the  Brownian motion $W_t$ are independent processes. As a result:
$$\mathbb{E}\left [\textnormal{exp} \biggl(\int_0^t \theta_s^c dC_s +\int_0^t \theta_{s_-}^J d J_s\right )\biggl | \mathcal{F}_t^\xi\biggl]=$$

\begin{equation}\label{4.61}\mathbb{E}\left [\exp \biggl(\int_0^t \theta_s^c (\mu_s-1/2\sigma_s^2)ds +\int_0^t \theta_s^c \sigma_s dW_s)\biggl) \biggl|\mathcal{F}_t^\xi\right]
 \mathbb{E}\left [\textnormal{exp}\biggl(\int_0^t \theta_{s_-}^J
 \log Z_{s_-} d N_s\biggl) \biggl| \mathcal{F}_t^\xi\right]. \end{equation}
 Let us calculate: $$\mathbb{E}\left [\textnormal{exp}\biggl(\int_0^t \theta_{s_-}^J \log Z_{s_-} d N_s\biggl) \biggl| \mathcal{F}_t^\xi\right].$$
Write $$\Gamma_t:=\exp\left(\int_0^t\alpha_{s_-}dN_s \right), \alpha_s=\theta_s^J \log Z_s.$$
Using the differentiation rule (see \cite{b11-1}) we obtain the following representation of $\Gamma_t$:
\begin{equation}\label{4.4-1}
\Gamma_t=\Gamma_0+M_t^J+\int_{]0,t]}\Gamma_s\int_{\mathbb{R}}(e^{\alpha_s}-1)\nu(dx)\lambda_s ds,
\end{equation}
where $$M_t^J=\int_{]0, t]}\Gamma_{s_-}(e^{\alpha_s}-1)dNs- \int_{]0,t]}\Gamma_s\int_{\mathbb{R}}(e^{\alpha_s}-1)\nu(dx)\lambda_s ds$$
is a martingale with respect to $\mathcal{F}_t^\xi$.
Using this fact and \eqref{4.4-1} we obtain:
\begin{equation}\label{4.62}
\mathbb{E}\left [\textnormal{exp}\biggl(\int_0^t \theta_{s}^J \log Z_{s_-}dN_s\biggl) \biggl| \mathcal{F}_t^\xi\right]=\exp\biggl(\int_0^t \lambda_s\biggl(\int_{\mathbb{R} }x^{\theta_s^J}\nu(dx)-1 \biggl) ds\biggl)
\end{equation}
We have from the differentiation rule:
\begin{equation}\label{4.62-1}
\mathbb{E}\left[e^ {u\int_0^t\sigma_s d W_s}\right]=
\exp \left \{\frac{1}{2}u^2\int_0^t \sigma_s^2 ds\right\}
\end{equation}
where $\sigma_t$ is the volatility of a market.
Substituting \eqref{4.62} and \eqref{4.62-1} into \eqref{4.61} we obtain:
$$
\mathbb{E}\left [\textnormal{exp} \biggl(\int_0^t \theta_s^c dC_s +\int_0^t \theta_{s_-}^J d J_s\right )\biggl | \mathcal{F}_t^\xi\biggl]=
$$

\begin{equation}\label{4.63}
\exp \biggl(\int_0^t \theta_s^c (\mu_s-1/2\sigma_s^2)ds +\frac{1}{2}\int_0^t (\theta_s^c \sigma_s)^2 ds )\biggl)
\exp\biggl(\int_0^t \lambda_s\biggl(\int_{\mathbb{R_+} }x^{\theta_s^J }\nu(dx)-1 \biggl) ds\biggl)
 \end{equation}
Substituting \eqref{4.63} to the expression for $L_t^{\theta^c,\theta^J}$ in \eqref{4.1} we have:

\begin{equation}\label{4.64}
L_t^{\theta^c,\theta^J}=\exp \biggl(\int_0^t \theta_s^c (\mu_s-1/2\sigma_s^2)ds +\int_0^t \theta_s^c \sigma_s dW_s)\biggl)\textnormal{exp}\biggl(\int_0^t \theta_{s_-}^J\log Z_{s_-} d N_s\biggl)\times
\end{equation}
$$
\biggl[\exp \biggl(\int_0^t \theta_s^c (\mu_s-1/2\sigma_s^2)ds +\frac{1}{2}\int_0^t (\theta_s^c \sigma_s)^2 ds )\biggl)
\exp\biggl(\int_0^t \lambda_s\biggl(\int_{\mathbb{R_+} }x^{\theta_s^J }\nu(dx)-1 \biggl) ds\biggl)\biggl]^{-1}=
$$
$$
\textnormal{exp}\left(\int_0^t \theta_s^c
\sigma_s dW_s -1/2 \int_0^t (\theta_s^c \sigma_s)^2
ds\right)\times
$$
$$\textnormal{exp}\left(\int_0^t \theta_{s_-}^J \log Z_{s_-}d N_s-\int_0^t \lambda_s\biggl(\int_{\mathbb{R_+}}x^{\theta_s^J }\nu(dx)-1\right)ds\biggl).
$$

If we present $L_t^{\theta^c,\theta^J}$ in the form  $L_t^{\theta^c,\theta^J}=e^{X_t}$ (see \eqref{4.64}) and and apply differentiation rule we obtain   SDE \eqref{4.60}. It follows from \eqref{4.60} that $L_t^{\theta^c,\theta^J}$ is a martingale.  $\square$

We shall derive the following condition for the discounted spot FX rate (\eqref{3.13}) to be martingale. These conditions will be used to calculate the risk-neutral  Esscher transform parameters $(\theta_t^{c, \ast})_{0\leq t\leq T}$, $(\theta_t^{J, \ast})_{0\leq t\leq T}$ and give to the measure $\textbf{Q}$. Then we shall use these values to find the no-arbitrage price of European call currency derivatives.

\textbf{Theorem 2.2.} \textit{Let the random Esscher transform be defined by \eqref{4.1}. Then
the martingale condition(for $S^d_t$, see \eqref{3.13}) holds if and only if Markov
modulated parameters ($\theta_t^c, \theta_t^J, 0\leq t\leq T$)
satisfy for all $0\leq t\leq T$ the condition:}
\begin{equation}\label{4.65}
r^f_t-r^d_t+\mu_t+\theta_t^c\sigma_t^2+\lambda_t^{\theta,
J }k_t^{\theta, J}=0
\end{equation}
\textit{where the random Esscher transform intensity $\lambda_t^{\theta,
J}$ of the Poisson Process and the main percentage jump size
$k_t^{\theta, J}$ are respectively given by}
\begin{equation}\label{4.66}
\lambda_t^{\theta, J}=\lambda_t \int_{\mathbb{R_+}}x^{\theta_s^J }\nu(dx),
\end{equation}
\begin{equation}\label{4.67}
 k_t^{\theta, J}=\frac{\int_{\mathbb{R_+}}x^{(\theta_t^J +1)}\nu(dx)}{\int_{\mathbb{R_+}}x^{\theta_t^J }\nu(dx)}-1.
\end{equation}
\textit{as long as}  $\int_{\mathbb{R_+}}x^{\theta_t^J +1}\nu(dx)<+\infty$.

\textbf{Proof of Theorem 2.2. } The martingale condition for the discounted spot FX rate $S^d_t$
\begin{equation}\label{4.11}
\mathbb{E}^{\theta^c, \theta^J}[S^d_t| \mathcal{H}_u]=S^d_u, \quad
t\geq u.
\end{equation}
To derive such a condition Bayes formula is used:
\begin{equation}\label{4.12}
  \mathbb{E}^{\theta^c,
\theta^J}[S^d_t|\mathcal{H}_u]=\frac{\mathbb{E}[L_t^{\theta^c,
\theta^J} S^d_t|\mathcal{H}_u]}{\mathbb{E}[L_t^{\theta^c,
\theta^J} |\mathcal{H}_u]},
\end{equation}
 taking into account that $L_t^{\theta^c,\theta^J}$ is a martingale with respect to $\mathcal{H}_u$, so:
\begin{equation}\label{4.13}
\mathbb{E}\biggl[L_t^{\theta^c, \theta^J}\biggl | \mathcal{H}_u\biggl]=L_u^{\theta^c,
\theta^J}.
\end{equation}

Using  formula \eqref{3.13} for the solution of the  SDE for the spot FX rate, we obtain an expression for the discounted  spot FX rate in the following form:

\begin{equation}\label{4.68}
S_t^d=S_u^d \exp\biggl(\int_u^t(r_s^f-r_s^d+\mu_s-1/2 \sigma_s^2) ds+\int_u^t \sigma_s dW_s+ \int_u^t \log Z_{s_-} d N_s\biggl), \quad t\geq u.
\end{equation}
Then, using \eqref{4.59}, \eqref{4.68} we can rewrite\eqref{4.13} in the following form:
\begin{equation}\label{4.69}
\mathbb{E}\biggl[\frac{L_t^{\theta^c, \theta^J}}{L_u^{\theta^c,
\theta^J}}S^d_t \biggl | \mathcal{H}_u\biggl]=S_u^d \;\mathbb{E}\biggl[\exp \left(\int_u^t \theta_s^c
\sigma_s dW_s -1/2 \int_0^t (\theta_s^c \sigma_s)^2
ds\right)\times
\end{equation}
$$\textnormal{exp}\left(\int_u^t \theta_{s_-}^J \log Z_{s_-}d N_s-\int_u^t \lambda_s\biggl(\int_{\mathbb{R_+}}x^{\theta_s^J}\nu(dx)-1\right)ds\biggl)\times $$
$$
\exp\biggl(\int_u^t(r_s^f-r_s^d+\mu_s-1/2 \sigma_s^2) ds+\int_u^t \sigma_s dW_s+ \int_u^t \log Z_{s_-} d N_s\biggl)| \mathcal{H}_u\biggl]=
$$
\begin{equation}\label{4.70}
 S_u^d \;\mathbb{E}\biggl[\exp \left(\int_u^t (\theta_s^c+1)
\sigma_s dW_s -1/2 \int_u^t ((\theta_s^c+1) \sigma_s)^2
ds\right)\times
\end{equation}
$$
\exp\biggl(\int_u^t(r_s^f-r_s^d+\mu_s+  \theta_s^c\sigma_s^2) ds\biggl)\;
\exp \biggl( \int_u^t \lambda_s\bigl(\int_{\mathbb{R}}e^{\theta_s^J x}\nu(dx)-1\bigl)ds\biggl)|\mathcal{H}_u\biggl]\times
$$
$$
\mathbb{E}\biggl[\exp \left(\int_u^t(\theta_s^c+1)
\log Z_{s_-}d N_s\right)| \mathcal{H}_u\biggl].
$$

Using expression for characteristic function of Brownian motion (see \eqref{4.62-1}) we obtain:
\begin{equation}\label{4.71}
 \mathbb{E}\biggl[\exp \left(\int_u^t (\theta_s^c+1)
\sigma_s dW_s -1/2 \int_u^t ((\theta_s^c+1) \sigma_s)^2
ds\right)| \mathcal{H}_u\biggl]=1.
\end{equation}

Using \eqref{4.62} we have:
\begin{equation}\label{4.72}
\mathbb{E}\biggl[\exp \left(\int_u^t(\theta_s^c+1)
\log Z_{s_-}d N_s\right)| \mathcal{H}_u\biggl]=\exp\biggl(\int_0^t \lambda_s\biggl(\int_{\mathbb{R_+} }x^{(\theta_s^J+1) }\nu(dx)-1 \biggl) ds\biggl).
\end{equation}
Substituting \eqref{4.71}, \eqref{4.72} into \eqref{4.70} we obtain finally:

\begin{equation}\label{4.73}
\mathbb{E}\biggl[\frac{L_t^{\theta^c, \theta^J}}{L_u^{\theta^c,
\theta^J}}S^d_t \biggl | \mathcal{H}_u\biggl]=S_u^d \exp\biggl(\int_u^t(r_s^f-r_s^d+\mu_s+  \theta_s^c\sigma_s^2) ds\biggl)\times
\end{equation}
$$
\exp \biggl( -\int_u^t \lambda_s\bigl(\int_{\mathbb{R_+}}x^{\theta_s^J}\nu(dx)-1\bigl)ds\biggl)\;\exp \biggl( \int_u^t \lambda_s\bigl(\int_{\mathbb{R_+}}x^{(\theta_s^J+1) }\nu(dx)-1\bigl)ds\biggl).
$$

From \eqref{4.73} we get the martingale condition for the discounted spot FX rate:

\begin{equation}\label{4.74}
r_t^f-r_t^d+\mu_t+  \theta_t^c\sigma_t^2+\lambda_t \biggl[\int_{\mathbb{R_+}}x^{(\theta_s^J+1) }\nu(dx)-\int_{\mathbb{R_+}}x^{\theta_s^J }\nu(dx)\biggl]=0.
\end{equation}
Prove now, that under the  Esscher transform the new Poisson process intensity and the  mean jump size are given by \eqref{4.66}, \eqref{4.67}.

Note that $L_t^J=\int_0^t \log Z_{s_-}dN_s$ is the jump part of L$\acute{\textnormal{e}}$vy process in the formula \eqref{3.12} for the solution of SDE for spot FX rate. We have:
\begin{equation}\label{4.75}
   \mathbb{E}_{\textbf{Q}} \left[ e^{L_t^J}\right]=\int_{\Omega}\exp\left(\int_0^t \log  Z_{s_-}dN_s\right)L_t^{\theta^{c},\theta^{J}}(\omega)d\textbf{P}(\omega),
\end{equation}
where $\textbf{P}$ is the initial  probability measure, $\textbf{Q}$ is a new risk-neutral measure.
Substituting the density of the Esscher transform \eqref{4.59} into \eqref{4.75} we have:
\begin{equation}\label{4.76}
 \mathbb{E}_{\textbf{Q}} \left[ e^{L_t^J}\right]=\mathbb{E}_\textbf{P}\biggl[\textnormal{exp}\left(\int_0^t \theta_s^c
\sigma_s dW_s -1/2 \int_0^t (\theta_s^c \sigma_s)^2
ds\right)-
\end{equation}
$$\int_0^t \lambda_s\biggl(\int_{\mathbb{R_+}}x^{\theta_s^J }\nu(dx)-1\biggl)ds\biggl) \biggl] \;\mathbb{E}_\textbf{P}\biggl[\textnormal{exp}\left(\int_0^t (\theta_{s}^J+1)\log Z_{s_-}d N_s\right)\biggl].
$$
Using \eqref{4.62} we obtain:
\begin{equation}\label{4.77}
\mathbb{E}_\textbf{P}\biggl[\textnormal{exp}\left(\int_0^t (\theta_{s}^J+1)\log Z_{s_-}d N_s\right)\biggl]=\exp\biggl(\int_0^t \lambda_s\biggl(\int_{\mathbb{R_+} }x^{(\theta_s^J +1)}\nu(dx)-1 \biggl) ds\biggl)
\end{equation}
Putting \eqref{4.77} to \eqref{4.76} and taking into account characteristic function of Brownian motion (see \eqref{4.62-1}) we have:
\begin{equation}\label{4.78}
 \mathbb{E}_{\textbf{Q}} \left[ e^{L_t^J}\right]=\exp\biggl(\int_0^t \lambda_s\biggl(\int_{\mathbb{R_+} }x^{\theta_s^J }\nu(dx)\left[\frac{\int_{\mathbb{R_+}}x^{(\theta_s^J +1)}\nu(dx)}{\int_{\mathbb{R_+}}x^{\theta_s^J }\nu(dx)}-1\right] \biggl) ds\biggl).
\end{equation}
Return to the initial measure $\textbf{P}$, but with different $\lambda_t^{\theta, J}, k_t^{\theta, J}$. We obtain:
\begin{equation}\label{4.79}
 \mathbb{E}_{\tilde{\lambda}, \tilde{\nu}} \left[ e^{L_t^J}\right]=\exp\biggl(\int_0^t \lambda_s^{\theta, J}\biggl(\int_{\mathbb{R_+}}x \tilde{\nu}(dx)-1\biggl)ds\biggl).
\end{equation}
Formula \eqref{4.66} for the new intensity $\lambda_t^{\theta, J}$ of the Poisson process follows directly from \eqref{4.78},\eqref{4.79}.
The new density of jumps $\tilde{\nu}$ is defined from \eqref{4.78}, \eqref{4.79} by the following formula:
\begin{equation}\label{4.80}
\frac{\int_{\mathbb{R_+}}x^{(\theta_t^J +1)}\nu(dx)}{\int_{\mathbb{R_+}}x^{\theta_t^J }\nu(dx)}=\int_{\mathbb{R_+}}x\tilde{\nu}(dx).
\end{equation}
Calculate now the  new mean jump size given jump arrival with respect to the new measure $\textbf{Q}$:
\begin{align}\label{4.81}
&k_t^{\theta, J}=\int_\Omega (Z(\omega)-1)d\tilde{\nu}(\omega)=\int_{\mathbb{R_+}} (x-1)\tilde{\nu}(dx)=\notag \\
&\int_{\mathbb{_+}} x\tilde{\nu}(dx)-1=
\frac{\int_{\mathbb{R_+}}x^{(\theta_t^J +1)}\nu(dx)}{\int_{\mathbb{R_+}}x^{\theta_t^J }\nu(dx)}-1.
\end{align}
So, we can rewrite martingale condition for the discounted spot FX rate in the  form in \eqref{4.65},
where $\lambda_t^{\theta, J}, k_t^{\theta, J}$ are given by \eqref{4.66}, \eqref{4.67} respectively. $\square$

Using \eqref{4.74} we have the following formulas for the families of the regime switching parameters satisfying the martingale condition \eqref{4.65}:
\begin{equation}\label{4.82}
 \theta_t^{c, \ast}=\frac{K_0+r_t^d-r_t^f-\mu_t}{\sigma_t^2},
\end{equation}
\begin{equation}\label{4.83}
 \theta_t^{J, \ast}: \int_{\mathbb{R_+}}x^{(\theta_t^{J, \ast} +1)}\nu(dx)-\int_{\mathbb{R_+}}x^{\theta_t^{J, \ast}} \nu(dx)=\frac{K_0}{\lambda_t},
\end{equation}
where $K_0$ is  any constant.
Note again, that the choice for these parameters is not unique.

In the next section we shall apply these formulas \eqref{4.82}, \eqref{4.83} to the  exponential distribution of jumps.

We now proceed to the general formulas for European calls (see \cite{b7}, \cite{b21}).
For the European call currency options with a strike price $K$ and the time of
expiration $T$ the price at time zero is given by:

\begin{equation}\label{4.31}
  \Pi_0(S, K, T, \xi)=\mathbb{E}^{\theta^{c,\ast}, \theta^{J,
  \ast}}\left[e^{-\int_0^T (r^d_s-r^f_s)ds}(S_T-K)^+\mid \mathcal{F}_t^\xi\right].
\end{equation}

 Let $J_i(t,T)$ denote the occupation time of $\xi$ in state $e_i$ over the period $[t, T], t<T$.  We  introduce several new quantities that will be used in future calculations:
\begin{equation}\label{4.32}
R_{t, T}=\frac{1}{T-t}\int_0^T (r_s^d-r_s^f)ds=\frac{1}{T-t}\sum_{i=1}^n (r_i^d-r_i^f)J_i(t, T),
\end{equation}
where $J_i(t, T):=\int_t^T <\xi_{s},\; e_i>ds$;
\begin{equation}\label{4.33}
U_{t, T}=\frac{1}{T-t}\int_t^T \sigma_s^2 ds=\frac{1}{T-t} \sum_{i=1}^n \sigma_i^2J_i(t, T);
\end{equation}
\begin{equation}\label{4.34}
 \lambda_{t, T}^{\theta^\ast J}=\frac{1}{T-t} \sum_{i=1}^n \lambda_i^{\theta^\ast J}J_i(t, T);
\end{equation}
\begin{equation}\label{4.35}
\lambda_{t, T}^{\theta^\ast }=\frac{1}{T-t}\int_t^T(1+k_s^{\theta^\ast J})\lambda_s^{\theta^\ast J}ds=\frac{1}{T-t}\sum_{i=1}^n (1+k_i^{\theta^\ast J})\lambda_i^{\theta^\ast J} J_i(t, T);
\end{equation}
\begin{equation}\label{4.36}
  V_{t, T, m}^2=U_{t, T}+\frac{m\sigma_J^2}{T-t},
\end{equation}
where $\sigma_J^2$ is the  variance of the distribution of the jumps.
\begin{equation}\label{4.37}
 R_{t, T, m}=R_{t,T}-\frac{1}{T-t}\int_t^T \lambda_s^{\theta^\ast J}k_s^{\theta^\ast J}ds+\frac{1}{T-t}\int_0^T\frac{\log(1+k_s^{\theta^\ast J})}{T-t}ds=
\end{equation}
$$
R_{t, T}-\frac{1}{T-t}\sum_{i=1}^n \lambda_i^{\theta^\ast J}k_i^{\theta^\ast J}+
\frac{m}{T-t}\sum_{i=1}^n \frac{\log(1+k_i^{\theta^\ast J})}{T-t}J_i(t,T),
$$
where $m$ is the number of jumps in the interval $[t, T]$, $n$ is the number of states of the Markov chain $\xi$.

From the pricing formula in Merton (1976, \cite{b21}) let us define (see \cite{b7})
\begin{equation}\label{4.38}
  \overline{\Pi_0}(S, K,T; R_{0, T}, U_{0, T}, \lambda_{0, T}^{\theta^{\ast}})=\sum_{m=0}^{\infty}\frac{e^{-T\lambda_{0, T}^{\theta^\ast,
  J}}(T\lambda_{0, T}^{\theta^\ast})^m}{m!}\times
\end{equation}
$$BS_0(S, K, T, V^2_{0, T, m}, R_{0, T,
  m})$$
where $BS_0(S, K, T, V^2_{0, T, m}, R_{0, T,
  m})$ is the standard Black-Scholes price formula (see \cite{b6})
with initial spot FX rate $S$, strike price $K$, risk-free rate $r$,
volatility square $\sigma^2$ and time $T$ to maturity.

Then, the European style call option pricing formula takes the form (see \cite{b7}):
\begin{equation}\label{4.39}
\Pi_0(S, K, T)=\int_{[0, t]^n}\overline{\Pi_0}(S, K,T; R_{0, T},
U_{0, T}, \lambda_{0, T}^{\theta^{\ast,
  J}})\times
\end{equation}
$$\psi(J_1, J_2,..., J_n)  dJ_1...dJ_n,$$
where $\psi(J_1, J_2,..., J_n)$ is the joint probability distribution
density for the occupation time, which is determined by the following characteristic function (See \cite{b11-2}):
\begin{equation}\label{4.39-1}
\mathbb{E} \left[\exp \bigl\{\langle u, J(t,T)\rangle \bigl\}\right]=\langle\exp \{(\Pi + diag(u)) (T-t)\}\cdot\mathbb{E}[\xi_0], \mathfrak{1}\rangle,
\end{equation}
where $\mathfrak{1}\in \mathbb{R}^n$ is a vector of ones, $u=(u_1,..., u_n)$ is a vector of transform variables, $J(t,T):=\{J_1(t,T),..., J_n(t,T)\}$.

\section{Currency option pricing  for exponential processes}

Because of  the restriction $Z_{s_-}>0$ we can not consider a double-exponential distribution of jumps (see \cite{b17}, \cite{b19}) in $\int_0^t \log Z_{s_-}dN_s $. Let us consider exponential distribution instead. It is defined by the following formula of density function:
\begin{equation}\label{4.84}
 \nu(x)=\theta e^{-\theta x}\biggl|_{x \geq 0}
\end{equation}
The mean value of this distribution is:
\begin{equation}\label{4.85}
  \textnormal{mean}(\theta)=\frac{1}{\theta}
\end{equation}
The variance of this distribution is:
\begin{equation}\label{4.86}
\textnormal{var}(\theta)=\frac{1}{\theta^2}
\end{equation}
 The exponential distribution like the double-exponential distribution has also memorylessness property.

 Let us derive the martingale condition and formulas for the regime-switching Esscher transform parameters in case of jumps driven by  the exponential distribution. Using the martingale condition for discounted spot FX rate \eqref{4.74} we obtain:

 \begin{equation}\label{4.87}
r_t^f-r_t^d+\mu_t+  \theta_t^c\sigma_t^2+\lambda_t \biggl[\frac{\Gamma(\theta_t^J+2)}{\theta^{\theta_t^J+1}}-\frac{\Gamma(\theta_t^J+1)}{\theta^{\theta_t^J}}\biggl]=0,
\end{equation}
 where we have such a restriction(and in the sequel): $\quad \theta_t^J>-1$.

Using \eqref{4.66}, \eqref{4.67} the random Esscher transform intensity $\lambda_t^{\theta,
J}$ of the Poisson Process and the main percentage jump size
$k_t^{\theta, J}$ are respectively given by
\begin{equation}\label{4.88}
\lambda_t^{\theta, J}=\lambda_t \frac{\Gamma(\theta_t^J+1)}{\theta^{\theta_t^J}},
\end{equation}
\begin{equation}\label{4.89}
 k_t^{\theta, J}=\frac{\theta_t^J+1}{\theta}-1.
\end{equation}

Using \eqref{4.83} we have the following formula for the families of regime switching Esscher transform  parameters satisfying martingale condition \eqref{4.87}:
\begin{equation}\label{4.90}
 \theta_t^{J, \ast}: \biggl[\frac{\Gamma(\theta_t^{J,\ast}+2)}{\theta^{\theta_t^{J,\ast}+1}}-\frac{\Gamma(\theta_t^{J,\ast}+1)}{\theta^{\theta_t^{J,\ast}}}\biggl]=\frac{K_0}{\lambda_t}.
\end{equation}
Let us simplify \eqref{4.90}:
\begin{equation}\label{4.91}
  \theta_t^{J, \ast}: \frac{\Gamma(\theta_t^{J,\ast}+1)}{\theta^{\theta_t^{J,\ast}}}\biggl(\frac{\theta_t^{J,\ast}+1}{\theta}-1\biggl)=\frac{K_0}{\lambda_t}.
\end{equation}
The formula for $\theta_t^{c, \ast}$ in this case is the same as in \eqref{4.82}.

With respect to to such values of the regime switching Esscher transform  parameters we have from \eqref{4.88}, \eqref{4.89}, \eqref{4.91}:
\begin{equation}\label{4.92}
  k_t^{J, \ast}=K_0/\lambda_t^{ J, \ast}.
\end{equation}

When we proceed to a new risk-neutral measure $\textbf{Q}$ we have the new $\tilde{\theta}$ in \eqref{4.84}. Using \eqref{4.80} we obtain:

\begin{equation}\label{4.93}
\tilde{\theta}=\frac{\theta}{\theta_t^J+1}.
\end{equation}
From \eqref{4.93} we arrive at interesting conclusion:  $\tilde{\theta}$ depends on time $t$. So, now the distribution of jumps changes depending on time(it is not the case before for the log double-exponential  distribution, where $\tilde{\theta}$ was actually a constant, see \cite{b7}). So, the compound Poisson Process depends not only on a number of jumps, but on moments of time when they arrive in this case. The same statement  is true for the mean jump size in \eqref{4.89}. But the  pricing formulas \eqref{4.31}-\eqref{4.39} are applicable to this case as well.

In the numerical simulations, we  assume that
the hidden Markov chain has three states: up, down, side-way, and the corresponding rate matrix is calculated using real Forex data for the thirteen-year period: from January 3, 2000 to November 2013. To calculate all probabilities we use the Matlab script (see the Appendix).

\section{Numerical simulations}

In the following   figures  we shall provide numerical simulations for the  case when amplitude of jumps is described by the exponential distribution. These plots show the dependence of a European-call option price on  $S/K$, where $S$ is the initial spot FX rate ($S=1$ in our simulations)), $K$ is a strike FX rate for various maturity times $T:$  0.5, 1, 1.5 in years and various values of a parameter $\theta:$ 2.5, 3.5, 5 in the exponential distribution.   Blue line stands for the exponential distribution of jumps, red-line is  for the dynamics without jumps. From these plots  we can make a conclusion that it is important to incorporate a jump risk into the spot FX rate models (described by the Black-Scholes equation without jumps red line on a plot is  below the blue  line standing for the exponential  distributions of jumps).

\begin{figure}[h!]
\begin{minipage}[h!]{0.3\linewidth}
\center{\includegraphics[width=0.8\linewidth]{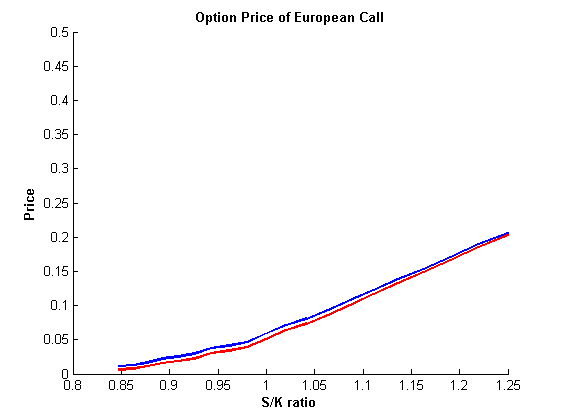} \\ $T=0.5, \;\theta=5$}
\end{minipage}
\hfill
\begin{minipage}[h!]{0.3\linewidth}
\center{\includegraphics[width=0.8\linewidth]{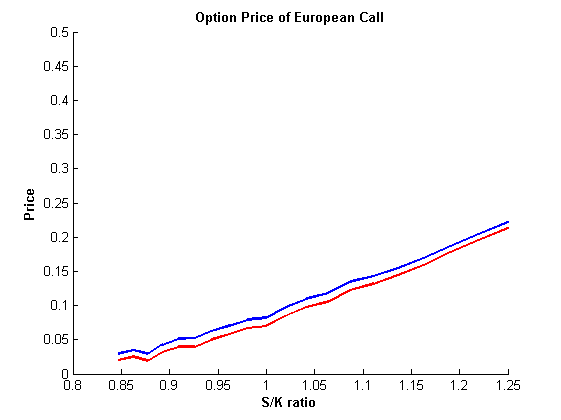} \\ $T=1.0, \;\theta=5$}
\end{minipage}
\hfill
\begin{minipage}[h!]{0.3\linewidth}
\center{\includegraphics[width=0.8\linewidth]{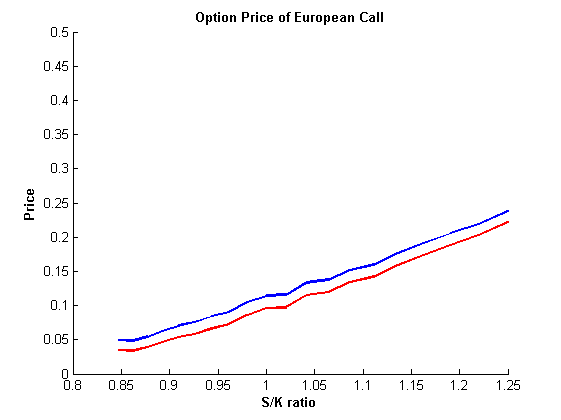} \\ $T=1.5, \;\theta=5$}
\end{minipage}
\caption{Option price of European Call: $\theta=5$}
\end{figure}
\begin{figure}[h!]
\begin{minipage}[h!]{0.3\linewidth}
\center{\includegraphics[width=0.8\linewidth]{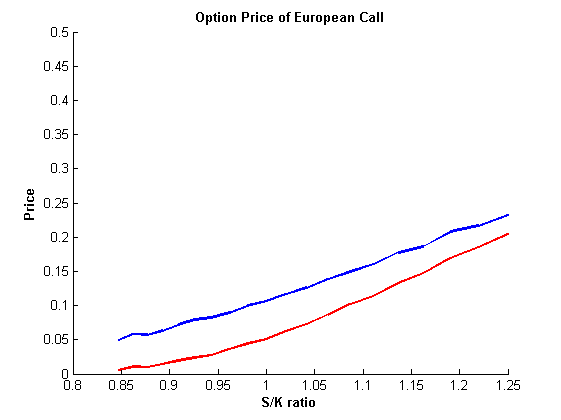} \\ $T=0.5, \;\theta=3.5$}
\end{minipage}
\hfill
\begin{minipage}[h!]{0.3\linewidth}
\center{\includegraphics[width=0.8\linewidth]{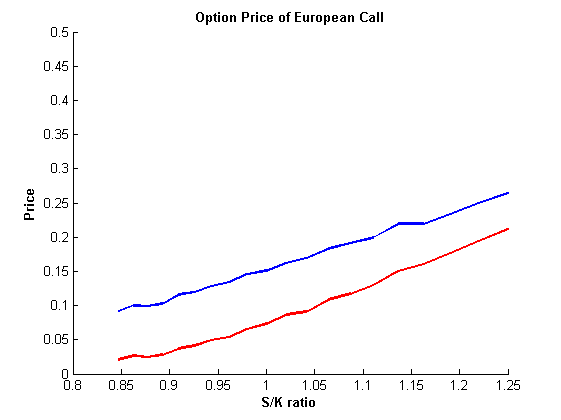} \\ $T=1.0, \;\theta=3.5$}
\end{minipage}
\hfill
\begin{minipage}[h!]{0.3\linewidth}
\center{\includegraphics[width=0.8\linewidth]{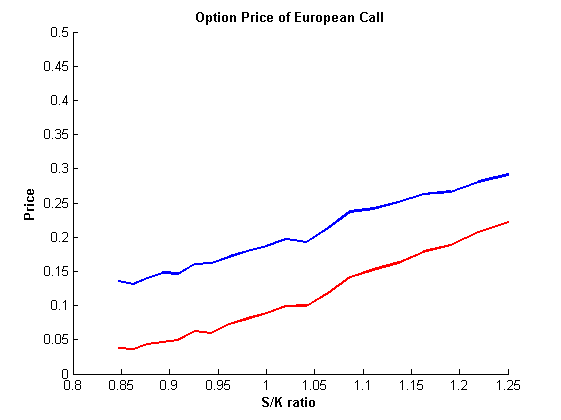} \\ $T=1.5, \;\theta=3.5$}
\end{minipage}
\caption{Option price of European Call: $\theta=3.5$}
\end{figure}

\begin{figure}[h!]
\begin{minipage}[H]{0.3\linewidth}
\center{\includegraphics[width=0.8\linewidth]{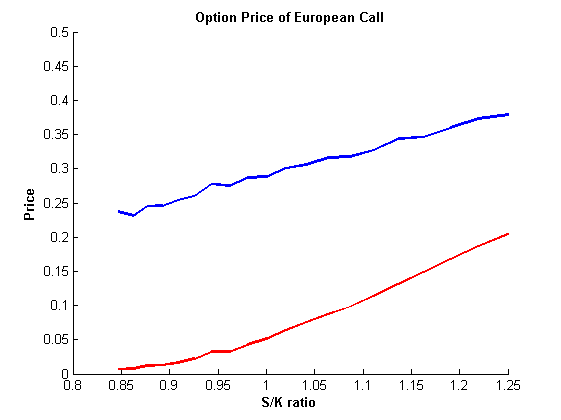} \\ $T=0.5, \;\theta=2.5$}
\end{minipage}
\hfill
\begin{minipage}[H]{0.3\linewidth}
\center{\includegraphics[width=0.8\linewidth]{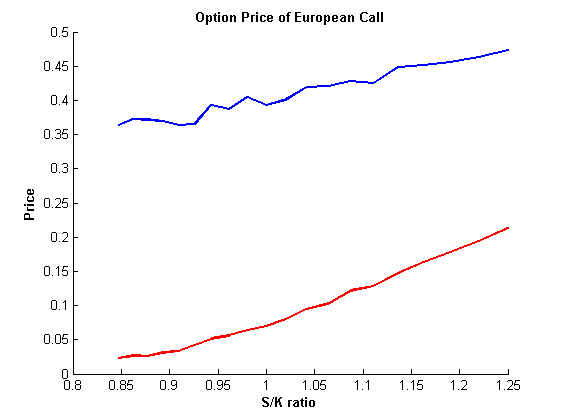} \\ $T=1.0, \;\theta=2.5$}
\end{minipage}
\hfill
\begin{minipage}[H]{0.3\linewidth}
\center{\includegraphics[width=0.8\linewidth]{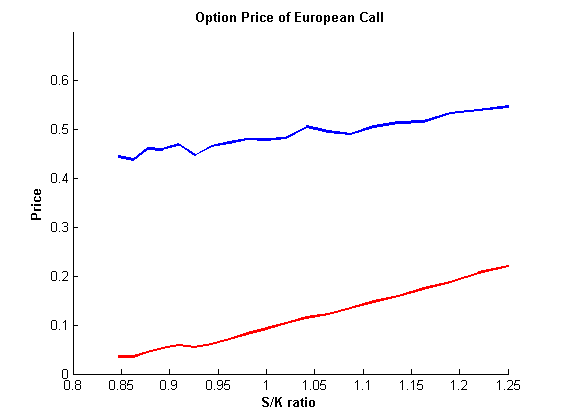} \\ $T=1.5, \;\theta=2.5$}
\end{minipage}
\caption{Option price of European Call: $\theta=2.5$}
\end{figure}

\section*{Appendix}
\textbf{The Matlab function used to calculate probability matrix for the Markov chain modeling  cross rates of currency pairs in the Forex market.}

We assume that the Markov chain has only three states: "trend up", "trend down", "trend sideway". Such a choice of states is justified by numerous articles for the FX market (See www.mql5.com).
In a file $\textnormal{MaxDataFile open.CSV}$ there are open prices of EURO/ESD currency pairs of Japanese  candles over a 13 year period. This file was generated in the platform MT5 using MQL5 programming language.

\begin{verbatim}
function [ Probab_matrix ] =Probab_matrix_calc1(candles_back_up, candles_back_down,
delta_back_up, delta_back_down, candles_up,candles_down, delta_up, delta_down )
Probab_matrix=zeros(3,3);
m_open=csvread('MaxDataFile_open.CSV');
[size_open temp]=size(m_open);
m_before=zeros(1,size_open);
upper_border=size_open-max(candles_up, candles_down);
delta_up=delta_up/10000;
delta_down=delta_down/10000;
count_up=0;
count_down=0;
count_sideway=0;
beforeborder=max(candles_back_up, candles_back_down)+1;
for i=beforeborder:size_open
    if (m_open(i)-m_open(i-candles_back_up)>=delta_up)
        m_before(i)=1;
    end
     if (m_open(i-candles_back_down)-m_open(i)>=delta_down)
        m_before(i)=-1;
     end
end;
for i=1:upper_border
    if(m_before(i)==1)
        if(m_open(i+candles_up)-m_open(i)>=delta_up)
        Probab_matrix(1,1)= Probab_matrix(1,1)+1;
        else
            if(m_open(i)-m_open(i+candles_down)>=delta_down)
                Probab_matrix(1,2)= Probab_matrix(1,2)+1;
            else
                Probab_matrix(1,3)= Probab_matrix(1,3)+1;
            end
        end
    end
    if(m_before(i)==-1)
        if(m_open(i+candles_up)-m_open(i)>=delta_up)
        Probab_matrix(2,1)= Probab_matrix(2,1)+1;
        else
            if(m_open(i)-m_open(i+candles_down)>=delta_down)
                Probab_matrix(2,2)= Probab_matrix(2,2)+1;
            else
                Probab_matrix(2,3)= Probab_matrix(2,3)+1;
            end
        end
    end
    if(m_before(i)==0)
        if(m_open(i+candles_up)-m_open(i)>=delta_up)
        Probab_matrix(3,1)= Probab_matrix(3,1)+1;
        else
            if(m_open(i)-m_open(i+candles_down)>=delta_down)
                Probab_matrix(3,2)= Probab_matrix(3,2)+1;
            else
                Probab_matrix(3,3)= Probab_matrix(3,3)+1;
            end
        end
    end
end
count_up=sum(Probab_matrix(1,:));
count_down=sum(Probab_matrix(2,:));
count_sideway=sum(Probab_matrix(3,:));
for j=1:3
      Probab_matrix(1,j)= Probab_matrix(1,j)/count_up;
      Probab_matrix(2,j)= Probab_matrix(2,j)/count_down;
      Probab_matrix(3,j)= Probab_matrix(3,j)/count_sideway
end
end
\end{verbatim}

For example run in Matlab:
\begin{verbatim}
[ Probab_matrix ] = Probab_matrix_calc1(30, 30, 10, 10, 30, 30, 10, 10);
\end{verbatim}
Probability matrix is as follows:

$$
\left( {\begin{array}{cc}
    \textnormal{up} & \textnormal{down}\;\;\textnormal{sideway} \\
    0.4408 & 0.4527\;\; 0.1065\\
   0.4818 &0.4149 \;\; 0.1033\\
   0.4820 &0.4119\;\; 0.1061
  \end{array} } \right)
$$

\end{document}